\documentclass{PoS}

\title{Neutrino pion-production on a nucleon}

\ShortTitle{Neutrino pion-production}

\author{\speaker{Fred Myhrer} \\ %\thanks{A footnote may follow.}\\
        University of South Carolina, Columbia, SC 29208, USA\\
         \email{myhrer@physics.sc.edu}}

\abstract{Heavy baryon chiral perturbation theory ($\chi$PT), where the $\Delta$ resonance is included, is used  
in order to examine the axial charged-current component of the weak interaction process at low neutrino energies. 
At leading chiral order the Adler theorems, derived using PCAC, are satisfied.  
At next-to-leading chiral order this effective field theory goes beyond these theorems. 
I will show that $\chi$PT generates deviation from the PCAC predictions,  
which means that some neutrino-nucleon models that are used in evaluating neutrino nucleus scattering amplitudes,  
might need modifications. }

\FullConference{The 9th International workshop on Chiral Dynamics\\
		17-21 September 2018\\
		Durham, NC, USA}

\begin{document}

\section{Introduction} 

Neutrino beams generated in accelerator-based neutrino experiments have an energy spectrum 
which are typically in the range of a few hundred MeV to a few GeV.  
In order to accurately reconstruct the incoming neutrino energy spectrum, to be used in probing the properties 
of the enigmatic neutrinos, 
the outgoing muon is measured assuming that  quasi-elastic neutrino-nucleus charge-current 
scattering dominates the process.   
Nuclear corrections due to two-nucleon correlations and other nuclear effects are incorporated in the 
theoretical evaluations of the neutrino-nuclear reactions.  
The relevant neutrino reactions occurring in this energy regime are the purely leptonic charged-current 
scattering, the neutrino pion-production reactions and deep inelastic scattering. 
A theoretical understanding of these
processes is required~\cite{garvey2011} in order to accurately 
reconstruct the neutrino energy-distribution from these nuclear measurements. 
In the evaluations of these neutrino-nuclear reaction processes, 
neutrino-nucleon interaction models are the basic theoretical inputs,  
e.g., see a recent discussion in Ref.~\cite{Sato2018}.  
Due to the conservation of the vector current (CVC), the vector component of the weak-interaction models 
has been tested by     
electron and photo-nucleon and nuclear reactions  and we assume that this vector interaction is well known. 
However, the neutrino-nucleon axial interaction is less well known and is the focus of this investigation. 

In this paper we will discuss the  axial component of the neutrino pion-production from nucleons 
This process  is an important background reaction 
to the purely lepton-nucleon charged-current scattering. 
The pion-production reactions are manifest at a few hundred MeV where the $\Delta$-degrees of freedom are important. 
At low neutrino energies below the threshold for $\Delta$ production, a 
low-energy effective field theory (EFT) is applicable. 
We will use  heavy baryon chiral perturbation theory ($\chi$PT),  which is an EFT respecting the symmetries and 
the symmetry breaking patterns of QCD. 
$\chi$PT  assumes that 
we have an expansion in a typical  momentum $p$ of the order of the pion mass $m_\pi$ which is small 
relative to the chiral scale $\Lambda_\chi \approx m_N\approx 1$GeV,
i.e., $p/\Lambda_\chi \ll 1$.
In $\chi$PT  we treat the nucleon mass $m_N \ll m_\pi$ as  heavy. 
Furthermore, it is assumed that the leading chiral order (LO) terms dominate and that the next order contributions
give only smaller corrections to the LO results. 
We investigate the axial current structure predicted by $\chi$PT and we augment $\chi$PT by including  
the $\Delta$ degrees of freedom. 
At LO $\chi$PT 
satisfies the Adler theorems~\cite{Adler} which were derived using the  
partial conservation of the axial current (PCAC) expression, see also the review~\cite{Llewellyn}.  
At next-to-leading chiral order (NLO) we will show that this effective field theory generates 
deviation from PCAC as expected.  
It should be noted that for s-wave pion production at threshold,   
$\chi$PT has already been used by Bernard {\it et al.}~\cite{Bernard1994} to go beyond Adler's soft-pion theorems,  
and their expressions contain  deviations from PCAC. 
If these NLO contributions are of practical importance, then 
this implies that some neutrino-nucleon models might need  modifications. 
At next-to-next-to-leading order (NNLO) nucleon vector- and axial-vector form factors 
are generated in the $\chi$PT evaluation, and  the Goldberger-Treiman relation acquires a correction term.  
This is similar to what is observed in $\chi$PT approaches 
to other weak interaction processes at NNLO, e.g., Ref.~\cite{Ando2000}. 
A covariant ChPT evaluation of the  pion-production process 
at NNLO has recently been published~\cite{Yao2018} and the results were presented at this conference 
and will be summarized at the end of this paper. 

%%%%%%%%%%%%%%%%%%%%%%%%%%%%%%%%%%%%%%%%%%%%%%%%%%%%%%%%%%
\section{The $\chi$PT matrix elements} 
%%%%%%%%%%%%%%%%%%%%%%%%%%%%%%%%%%%%%%%%%%%%%%%%%%%%%%%%%%

The $\chi$PT lagrangian, ${\mathcal L}_\chi$   is written as a sum of terms where the leading order term contributes 
the dominant amplitude to the process under study. The next order terms of the chiral lagrangian 
ordinarily give smaller corrections to the LO amplitude. 
In this note we will evaluate the 
LO and NLO amplitudes which require only the following terms of the chiral lagrangian,  
\begin{eqnarray}
{\mathcal L}_\chi={\mathcal L}^{(1)}_{\pi N}+{\mathcal L}^{(2)}_{\pi N} 
+ {\mathcal L}^{(2)}_{\pi\pi}+\cdots \; ,  
%\nonumber
\end{eqnarray} 
where ${\mathcal L}^{(2)}_{\pi \pi} $ is the lowest order pion lagrangian. 
The leading order pion-nucleon $\chi$PT lagrangian, ${\mathcal L}^{(1)}_{\pi N} $, 
contains the nucleon velocity chosen to be 
$v^\mu=(1, \vec{0})$ and the nucleon spin $S^\mu=(0,\frac{1}{2}\vec{\sigma})$, 
\begin{eqnarray}
\mathcal{L}^{(1)}_{\pi N}&=& N^\dagger (iv\cdot {\cal D}+g_A S\cdot u) N \, , 
%\nonumber 
\end{eqnarray} 
where $g_A$ is the nucleon axial coupling constant and the expressions for ${\cal D}^\mu$ and $u^\mu$ 
can be found in, e.g.,  Ref.~\cite{bkm95}. 
The NLO correction terms are contained in ${\mathcal L}^{(2)}_{\pi N} $.  
The explicit expressions for the next order lagrangians,   
${\mathcal L}^{(2)}_{\pi N} $      %and ${\mathcal L}^{(2)}_{\pi \pi} $ can be found 
are given in, e.g., Ref.~\cite{bkm95}. 
%\vspace{15mm} 
It is assumed that the next-to-next-to-leading order corrections (NNLO) give smaller corrections than the NLO 
contributions. We will briefly discuss the importance of the NNLO contributions at the end of this article. 
We mention that the effective  
$\Delta$ degrees of freedom are an important part in understanding the low-energy neutrino pion productions. 
We  introduced the effective $\Delta$ resonance following Hemmert et al.~\cite{hemmert}, where it is assumed that 
the nucleon $\Delta$ mass difference, $m_\Delta - m_N = \delta_{N\Delta}$ is 
considered to be of the order of the pion mass, $m_\pi$. 
This differ from the Pascalutsa Phillips treatment of the $\Delta$ resonance~\cite{Pascalutsa} used in the 
relativistic ChPT approach of Yao {\it et al.}~\cite{Yao2018}.

\begin{figure}[t]
\begin{center}
\includegraphics[scale=0.6]{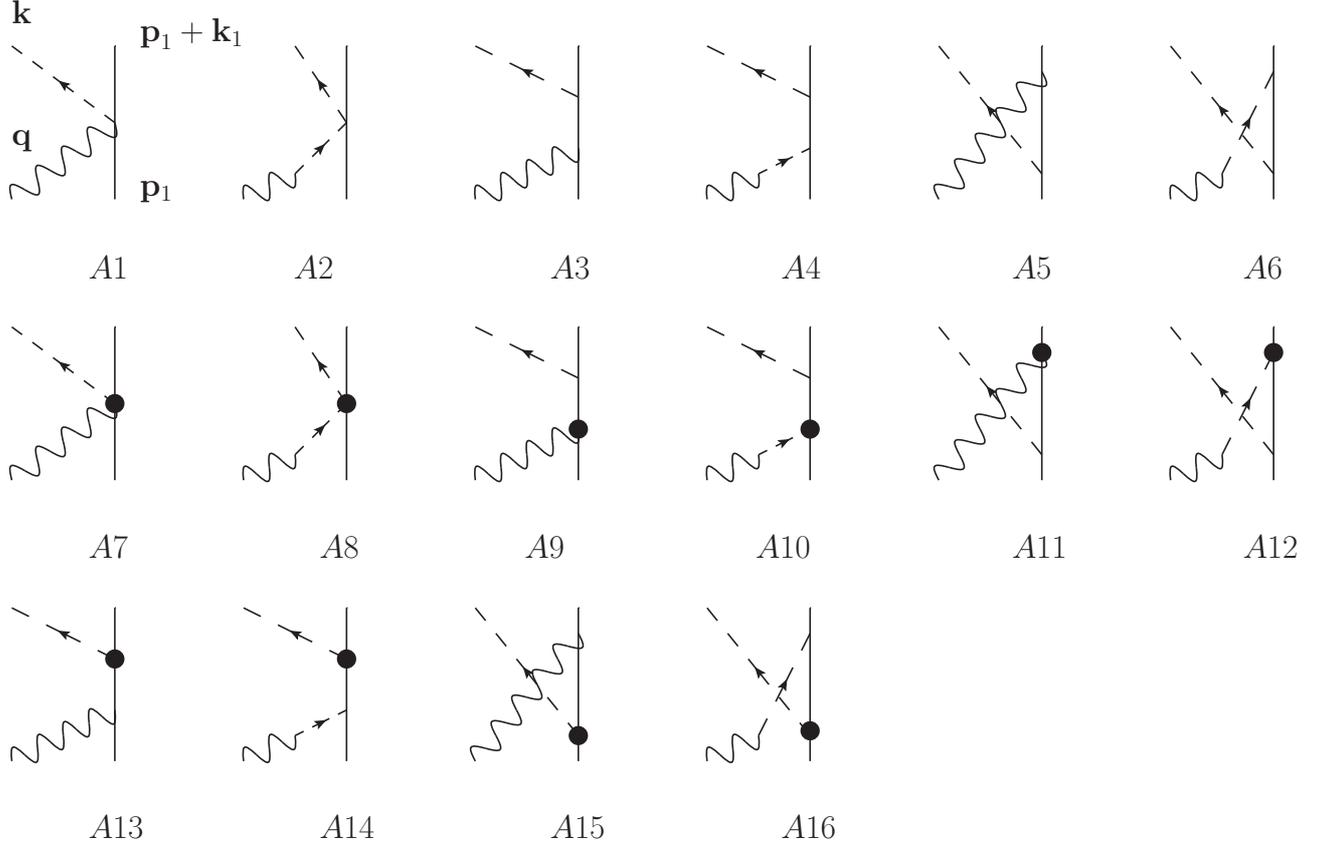}
\end{center}
\caption{ The incoming wavy line is the axial probe  
with isospin index $\alpha$ and four-momentum $q$. 
The incoming axial probe contains the pion-pole as shown explicitly in the diagrams.  
The dashed line is the outgoing pion 
of isospin index $\beta$ and four-momentum $k$. The solid line is the nucleon line with outgoing momentum 
$p_1^\prime= p_1+k_1$.  
The blobs signify vertices derived from ${\mathcal L}^{(2)}_{\pi N} $.  
When we evaluate the contributions from the effective $\Delta$, 
the internal nucleon lines, i.e., nucleon propagators, are replaced 
by $\Delta$ propagators .  }
\label{fig:feyn_diags} 
\end{figure}

%%%%%%%%%%%%%%%%%%%%%%%%%%%%%%%%%%%%%%%%%%%%%%%%%%%%%%%%%%
\subsection{A short note on the heavy nucleon (and $\Delta$) propagator} 
%%%%%%%%%%%%%%%%%%%%%%%%%%%%%%%%%%%%%%%%%%%%%%%%%%%%%%%%%%
 
 In this work the 
 nucleon and $\Delta$ propagators are treated differently from the usual HBChPT approach. 
%In Ref.~\cite{bkm95} 
In the usual HBChPT lagrangian  approach 
the  heavy nucleon propagator of momentum $p_1^\prime + k$ is  
derived from ${\mathcal L}^{(1)}_{\pi N} = N^\dagger \Big\{ v\cdot \partial + \cdots \Big\}N$, 
%e.g., Refs.~\cite{bkm95,ulf1},  
and is  
\begin{eqnarray}
\frac{i}{v\cdot (p_1^\prime + k) }\, . 
%\nonumber
\end{eqnarray} 
The next order lagrangian ${\mathcal L}^{(2)}_{\pi N} $
contains the kinetic Schr\"{o}dinger operator,   
\begin{eqnarray}
{\mathcal L}^{(2)}_{\pi N} &=& N^\dagger \left\{ -\frac{1}{2m_N} {\cal D}\cdot {\cal D} + \frac{1}{2m_N} (v\cdot {\cal D})^2
+\cdots  \right\}  N \, , 
%\nonumber 
\label{eq:L2}
\end{eqnarray} 
where ${\cal D}^\mu = \partial^\mu + \cdots$. 
The kinetic terms of 
${\mathcal L}^{(2)}_{\pi N} $ in Eq.~(\ref{eq:L2}) 
generate propagator corrections 
at second order, i.e., $1/(2m_N)$, as shown  by Bernard {\it et al.}~\cite{bkm95,ulf1}. 
In the  low-energy  pion production in two-nucleon collisions~\cite{baru}, 
it appears to be more effective to treat the   
nucleon propagator in the following manner, see Refs.~\cite{wirzba,sato} for discussions,   
\begin{eqnarray}
\frac{i}{v\cdot (p_1^\prime + k) - \frac{(\vec{p}_1^{\,\prime}+\vec{k})^2}{2m_N} } &=& \frac{i}{v\cdot k} 
\left\{ 1 + \frac{\vec{k}^{\, 2} + 2\vec{p}_1^{\,\prime}\cdot\vec{k} }{2m_N (v\cdot k)} +\cdots
\right\}  \, .
%\nonumber
\label{eq:prop}
\end{eqnarray} 
The essential argument is that since $(v\cdot p_1^\prime ) \approx \frac{ (\vec{p}_1^{\,\prime})^2}{2m_N} $,  
one of the leading terms is cancelled by a $1/(2m_N)$ term.      
We therefore expand the propagator in orders of $m_N^{-1}$ as in Eq.~(\ref{eq:prop}). 

%%%%%%%%%%%%%%%%%%%%%%%%%%%%%%%%%%%%%%%%%%%%%%%%%%%%%%%%%%
\subsection{The axial-current contribution to the matrix elements} 
%%%%%%%%%%%%%%%%%%%%%%%%%%%%%%%%%%%%%%%%%%%%%%%%%%%%%%%%%%
One remark concerns a leading order diagram, A2, in Fig.~\ref{fig:feyn_diags}. 
The LO  Weinberg-Tomozawa (W.-T.) vertex in diagram A2  contains 
the sum of the energies for the incoming and outgoing pions, $v\cdot (q+k)$, and  
four-momentum conservation gives, when expanded, the expression 
\begin{eqnarray}
v\cdot (q+k)&=& v\cdot k +v\cdot p^\prime_1-v\cdot p_1 +v\cdot k = 2\omega_\pi +
\frac{\vec{p}_1^{\ \prime 2} -\vec{p}_1^{\ 2}}{2m_N}+ \cdots \, . 
%\nonumber 
\label{eq:WT}
\end{eqnarray} 
Using the first term on the r.h.s. of the last equation~(\ref{eq:WT}), we
confirm  that the LO hadronic matrix elements reproduce  
the soft-pion theorem or PCAC. 
The expression, which follows below,  is derived from the Feynman diagrams in Fig.~\ref{fig:feyn_diags}.  
The nucleon axial-vector probe is indicated by the axial four-vector 
$\epsilon_A^\mu = (v\cdot \epsilon_A, \vec{\epsilon}_A)$. 
If PCAC is operative, we expect to find that each axial operator of the matrix elements has the 
following structure 
$\epsilon_A^\mu -\left(\frac{(\epsilon_A\cdot q)}{q^2-m_\pi^2}\right)l^\mu$,    
where $l^\mu$ is the $\mu$'th component of a four-vector. 
As is evident from the expression in Eq.~(\ref{eq:old}), we find some terms in the matrix element which 
do not conform to the expected PCAC-like expression. 
On closer inspections, these 
terms all originate from the $1/(2m_N)$ contributions of the LO diagram  A2 (see below) and 
from the NLO diagrams A7 and A8. 
These are the diagrams where the axial probe produce a pion in a single vertex 
on the nucleon. 
\begin{eqnarray}
M_A  \!\!\!\! &\simeq &\!\!\!\! 
 \epsilon^{\alpha\beta c}\tau_1^c\! \left[ \! \frac{(-1)}{2f_\pi}\!
\left( \!\!\epsilon_A\!\cdot\! v\! -\!\frac{ (\epsilon_A\!\cdot\! q) k\!\cdot\! v }{q^2\! -\! m_\pi^2}\!\right)\! 
+\! \frac{g_A^2}{2f_\pi\omega_\pi}\!\left(\! \!\vec{\epsilon_A}\!\cdot\!\vec{k}\! -\!
\frac{ (\epsilon_A\!\cdot\! q) \vec{q}\!\cdot\!\vec{k} }{q^2\! -\! m_\pi^2} \right)\!\!\right]\!\! -\! 
 \frac{\delta^{\alpha\beta} g_A^2}{2f_\pi\omega_\pi}\!
 \left(\!\! \vec{\epsilon_A}\!\cdot\! (\vec{\sigma}_1\!\!\times\!\vec{k})\! -\!  
\frac{ (\epsilon_A\!\cdot\!  q) \vec{q}\!\cdot\! (\vec{\sigma}_1\!\!\times\!\vec{k}) }{q^2\! -\! m_\pi^2}\!  \right) 
\nonumber \\ &&\!\!\!\!\! 
+ \! \frac{ \epsilon^{\alpha\beta c}\tau_1^c }{2f_\pi}\left[ 
\frac{ (\epsilon_A\!\cdot\! q)}{(q^2\! -\! m_\pi^2)} \right] 
\frac{  \left( \vec{p}_1^{\ \prime 2}\! -\!\vec{p}_1^{\ 2} \right) }{4m_N }  +
\nonumber \\ &+&\!\!\!\! 
\frac{ g_A^2 }{4m_N f_\pi ( v\!\cdot\! k\! +\! i\eta)}\! 
\left[ \vec{\epsilon_A}\!-\!\left(\!\frac{(\epsilon_A\!\cdot\! q) }{q^2\! -\! m_\pi^2 }\!\right) \vec{q} \right]\!\cdot\!
\left\{ \epsilon^{\alpha\beta c} \tau_1^c \left[ \vec{k} 
\Big[ \frac{\vec{k}\!\cdot\! (\vec{p}_1^{\,\prime}\!+\!\vec{p}_1)}{v\!\cdot\! k }\Big]
\! +\! i \Big(\vec{\sigma}_1\!\times\!\vec{k}\Big)
\Big[ \frac{\vec{k}\!\cdot\! (\vec{k}\! +\! \vec{k}_1)}{v\!\cdot\! k }    \Big] \right] \! 
\right. \nonumber \\ &&\!\!\!\!\left. 
-\!\delta^{\alpha\beta }\!\left[ \left(\vec{\sigma}_1\!\times\!\vec{k}\right) 
\Big[ \frac{\vec{k}\!\cdot\! (\vec{p}_1^{\,\prime}\!+\!\vec{p}_1)}{v\!\cdot\! k }\Big]\!
- i \vec{k} 
\Big[ \frac{\vec{k}\!\cdot\! (\vec{k}\! +\! \vec{k}_1)}{v\!\cdot\! k }    \Big] 
\right] \right\}   
\nonumber \\ &+&\!\!\!\! 
(-i) \frac{\delta^{\alpha\beta }}{f_\pi} \left\{ \left[ \epsilon_A\! -\!\left(\!\frac{(\epsilon_A\!\cdot\! q) }{q^2\! -\! m_\pi^2}\!\right) 
q\right] \!\cdot\!
\left[  
v\, \left(2 c_2\! -\!\frac{g_A^2}{4m_N}\right)\! (v\!\cdot\! k)\!  +\!  k\, 2 c_3\!  \right] 
\nonumber \right. \\  && \left. 
+ \left(\frac{\epsilon_A\cdot q}{q^2-m_\pi^2}\right)  4 c_1 m_\pi^2 
\right\} 
%\nonumber \\ & + &\!\!\!\! 
+\frac{ \epsilon^{\alpha\beta c}\tau_1^c}{f_\pi} \left\{ \! 
 \left(\frac{1\! +\!\kappa_V}{4m_N }\right) \!
 i\vec{\epsilon}_A\!\cdot\!\left(\vec{\sigma}_1\!\times\! \vec{q}\right)\! +\! 
 \frac{\vec{\epsilon}_A\!\cdot\! (\vec{p}_1\! +\!\vec{p}_1^{\,\prime}) }{8m_N }   +\! 
 \right. 
 \nonumber \\ &+&\!\!\!\! \left. 
 \left[ \vec{\epsilon}_A\!-\! \left(\!\frac{(\epsilon_A\!\cdot\! q) }{q^2\! -\! m_\pi^2} \!\right)(\vec{q}\! +\!\vec{k})   \right]\!\cdot\! 
 \left[ \frac{  (\vec{p}_1\! +\!\vec{p}_1^{\,\prime}) }{8m_N } \! 
-\! \left(  c_4 +\frac{1}{4m_N}\right) 
i  \left(\vec{\sigma}_1\!\times\! \vec{k}\right) \right]  
\right\} 
\nonumber \\ &+&\!\!\!\! \!
\frac{(-i) g_A^2}{4 m_N f_\pi}\! \left(\!\frac{ 1 }{v\!\cdot\! k\! +\! i\eta}\!\right) \!\! 
\left[\! \epsilon_A\! -\!  
\frac{(\epsilon_A\!\cdot\! q)  q}{q^2-m_\pi^2}\!\right]\!\cdot\! v \!
\left\{ \! \delta^{\alpha\beta }\!\left[ \! 
\vec{k}^{\, 2}\!\! +\! i\vec{\sigma}_1\!\cdot \!\! 
\left(\vec{k}\!\times\! (\vec{p}_1\! +\!\vec{p}_1^{\ \prime})\!\right) \! 
%\right.\right. \nonumber \\ &+&\!\!\!\!\!\!\left. \left. 
\!\right]\!\! 
+\! i\epsilon^{\beta\alpha c}\tau_1^c\! \left[  \vec{k}\!\cdot\! (\vec{p}_1\!\! +\!\vec{p}_1^{\ \prime}) \right]\!  
\right\} 
\nonumber \\ &+& \!\!\!\! \!
\frac{(-i) g_A^2}{4 m_N f_\pi} \!\left(\frac{\omega_\pi}{v\cdot k\! +\!i\eta}\right)\!\!
\left[\vec{\epsilon}_A\! -\!
\left(\frac{\epsilon_A\cdot q}{q^2-m_\pi^2}\right)\! \vec{q}\ \right] \! \cdot \! 
\left\{ \delta^{\alpha\beta }\! \left[ 
(\vec{p}_1^{\ \prime}\! -\! \vec{p}_1\! +\!\vec{k}) \! 
+\! i 
\left(\vec{\sigma}_1\!\times\! (\vec{p}_1\! +\!\vec{p}_1^{\ \prime})\right) 
\right] 
\right.\nonumber \\ &+& \left. \!\!\!\! 
i\epsilon^{\beta\alpha c}\tau_1^c\!\! \left[ 
(\vec{p}_1\! +\!\vec{p}_1^{\ \prime})\! 
+\! i 
\left(\vec{\sigma}_1\!\times\! (\vec{p}_1^{\ \prime}\! -\! \vec{p}_1\!+\! \vec{k} )\right) 
\right] 
\right\} 
\label{eq:old}
%\nonumber 
\end{eqnarray}  
In this expressions the low-energy constants (LECs), $c_i$ where $i=1, \cdots ,4$, 
are determined by pion-nucleon scattering data. 
The first line of  Eq.~(\ref{eq:old}) contains the LO contributions which exhibit the expected PCAC structure. 
The  second line in Eq.~(\ref{eq:old}) contains the $m_N^{-1}$ correction of the W.-T. vertex 
given in Eq.~(\ref{eq:WT}) while 
the third and fourth lines are the  $m_N^{-1}$ correction   
from the nucleon propagators  of the LO diagrams in Fig.~\ref{fig:feyn_diags}.  
The main deviations from the expected PCAC-like expressions are the 
three terms in the sixth line.   
The most conspicuous  one contains  the nucleon iso-vector magnetic moment $1+\kappa_V$ .  
In some models of this neutrino reaction, this contribution could be associated with the ``heavy" $\rho$-meson 
which couple to the nucleon in a $t$-channel-like exchange, e.g, Ref.~\cite{Sato2018}.   
%
%Are these NLO terms included in the  low-energy limit of  
In models for the elementary neutrino - nucleon reaction  amplitudes   
used in neutrino nucleus descriptions, it is frequently assumed that PCAC is valid. 
However, some of the NLO contributions in Eq.~(\ref{eq:old}), which are derived using $\chi$PT, 
reveal that PCAC should be corrected at NLO (and NNLO), corrections which  
were included in Ref.~\cite{Bernard1994}. 
Furthermore, we  
observe that the term with the LEC, $c_1$, which is  connected to the pion-nucleon sigma term,  only contributes 
to the pion-pole amplitude of the axial current.

The tree-diagrams involving an intermediate $\Delta$ excitation are included and at  
leading order (as well as the next order) the $\Delta$ contribution 
to the hadronic matrix element 
has the structure   expected from PCAC, e.g., at the lowest order we obtain  
\begin{eqnarray}
{\cal M}_\Delta\!\!\!\! &=&\!\!\!\! 
\left( (-i) \frac{g^2_{\pi N\Delta}}{f_\pi} \right) 
%u^\dagger(p_1^\prime) 
\left\{  
\frac{-4}{9}\delta^{\alpha\beta} 
\left[ \vec{\epsilon}_A\!\cdot\!\vec{k} -\frac{(\epsilon_A\cdot q) (\vec{q}\cdot\vec{k}\, ) }{q^2-m_\pi^2}  \right] 
\left[ \frac{2\delta_{N\Delta} }{(v\cdot k)^2- (\delta_{N\Delta})^2}\right] 
\right.
\nonumber \\ &+& \left. 
 \frac{+2i}{9} \epsilon^{\alpha\beta c}\tau_1^c
\left[ \vec{\epsilon}_A\!\cdot\!\vec{k} -\frac{(\epsilon_A\cdot q) (\vec{q}\cdot\vec{k}\, )}{q^2-m_\pi^2} 
\right] \left[ \frac{2(v\cdot k)}{(v\cdot k)^2 - (\delta_{N\Delta})^2}\right]   
\right.
\nonumber \\ &+& \left. 
\frac{2i}{9}\delta^{\alpha\beta}
\left[ \vec{\epsilon}_A\!\cdot\!(\vec{\sigma}_1\times \vec{k}) - 
\frac{(\epsilon_A\cdot q) \vec{\sigma}_1\cdot (\vec{k}\!\times\!\vec{q}\, ) }{q^2-m_\pi^2} \right] 
\left[ \frac{2(v\cdot k)}{(v\cdot k)^2 - (\delta_{N\Delta})^2}\right] 
\right.
\nonumber \\ &+& \left. 
 \frac{(+1)}{9} \epsilon^{\alpha\beta c}\tau_1^c 
 \left[   \vec{\epsilon}_A\!\cdot\!(\vec{\sigma}_1\times \vec{k}) - 
\frac{ (\epsilon_A\cdot q) \vec{\sigma}_1\cdot (\vec{k}\!\times\!\vec{q}\, )}{q^2-m_\pi^2} \right] 
\left[ \frac{2\delta_{N\Delta} }{(v\cdot k)^2-  (\delta_{N\Delta})^2} \right] 
\right\}   %u(p_1) 
%\nonumber 
\end{eqnarray}

%%%%%%%%%%%%%%%%%%%%%%%%%%%%%%%%%%%%%%%%%%%%%%%%%%%%%%%%%%
\section{A short note on the NNLO contributions } 
%%%%%%%%%%%%%%%%%%%%%%%%%%%%%%%%%%%%%%%%%%%%%%%%%%%%%%%%%% 
The neutrino production of pions at low energy has been evaluated by Yao {\it et al.}~\cite{Yao2018}  in  
a covariant ChPT approach. Yao {\it et al.} included LO, NLO and  NNLO contributions of this weak-interaction 
reaction. 
As shown in Ref.~\cite{Yao2018}, the NNLO corrections are larger than expected and are very important in order to 
reproduce the measured low-energy neutrino cross sections. 
It is expected  that  the $\Delta$ contributions are dominant for neutrino energies of about 1 GeV,  
and the $\Delta$ resonance is included in models which evaluate the neutrino-nuclear reactions, see, e.g., 
Ref.~\cite{HNV} and later works along these lines~\cite{Watson,HN}. 
In addition, Ref.~\cite{HNV}  included the leading order ChPT contributions.
Yao {\it et al.} showed that at 
neutrino energies below the $\Delta$-resonance, this resonance is  important for 
the final proton $\pi^+$ final state. However, at a neutrino energy of about 400 MeV and 
for  final neutron $\pi^+$ or proton $\pi^0$ states, their covariant ChPT evaluation  showed that 
the nucleon degrees of freedom are 
very important, and that the $\Delta$ is not as prominent as assumed in many model calculations.

%%%%%%%%%%%%%%%%%%%%%%%%%%%%%%%%%%%%%%%%%%%%%%%%%%%%%%%%%%
\section{Summary } 
%%%%%%%%%%%%%%%%%%%%%%%%%%%%%%%%%%%%%%%%%%%%%%%%%%%%%%%%%% 

We use the effective low-energy field theory called  heavy baryon chiral perturbation theory and 
reproduce the soft-pion theorems at leading order. 
We show at next-to-leading order, which goes beyond the soft-pion theorems, that 
the axial current contribution 
to the neutrino pion-production reaction on a nucleon gives corrections to PCAC.  
In other words, 
our finding does not conform with some model descriptions which assumes PCAC is valid,  and which are 
used in neutrino-nucleus reactions.

\section{Acknowledgement} 
Part of this work was done during the summers spent at the Ruhr University Bochum, where the author enjoyed 
many useful and constructive discussions. 
The author is grateful for numerous discussions on this topic  with Prof. Toru Sato. 

%%%%%%%%%%%%%%%%%%%%%%%%%%%%%%%%%%%%%%%%%%%%%%%%%%%%%% 
%%%%%%%%%%%%%%%%%%%%%%%%%%%%%%%%%%%%%%%%%%%%%%%%%%%%%% 


\begin{thebibliography}{99} 

\bibitem{garvey2011} 
%\cite{Gallagher:2011zza}
%\bibitem{Gallagher:2011zza} 
  H.~Gallagher, G.~Garvey and G.~P.~Zeller,
  `{\it Neutrino-nucleus interactions}, 
  Ann.\ Rev.\ Nucl.\ Part.\ Sci.\  {\bf 61}, 355 (2011).
  %doi:10.1146/annurev-nucl-102010-130255
  %%CITATION = doi:10.1146/annurev-nucl-102010-130255;%%
  %47 citations counted in INSpIRE as of 13 Feb 2019
%G. Garvey, presentation at Fermi National Laboratory (2011), \\
%http://minerva-docdb.fnal.gov/cgi-bin/ShowDocument?docid=5878 

\bibitem{Sato2018} 
%\cite{Sobczyk:2018ghy}
%\bibitem{Sobczyk:2018ghy} 
  J.~E.~Sobczyk, E.~Hern\'{a}ndez, S.~X.~Nakamura, J.~Nieves and T.~Sato,
  {\it Angular distributions in electroweak pion production off nucleons: odd parity hadron terms, 
  strong relative phases and model dependence}, 
  Phys.\ Rev.\ D {\bf 98}, 073001 (2018)
 % doi:10.1103/physRevD.98.073001
  [arXiv:1807.11281 [hep-ph]].
  %%CITATION = doi:10.1103/physRevD.98.073001;%%
  %1 citations counted in INSpIRE as of 13 Feb 2019 

\bibitem{Adler}
%\cite{Adler:1968tw}
%\bibitem{Adler:1968tw} 
  S.~L.~Adler,
  {\it photoproduction, electroproduction and weak single pion production in the (3,3) resonance region}, 
  Annals Phys.\  {\bf 50}, 189 (1968).
 % doi:10.1016/0003-4916(68)90278-9 

\bibitem{Llewellyn} 
%\cite{LlewellynSmith:1971uhs}
%\bibitem{LlewellynSmith:1971uhs} 
  C.~H.~Llewellyn Smith,
  {\it Neutrino Reactions at Accelerator Energies}, 
  Phys.\ Rept.\  {\bf 3}, 261 (1972).
  %doi:10.1016/0370-1573(72)90010-5
  %%CITATION = doi:10.1016/0370-1573(72)90010-5;%%
  %874 citations counted in INSpIRE as of 13 Feb 2019   

\bibitem{Bernard1994} 
%\cite{Bernard:1993xh}
%\bibitem{Bernard:1993xh} 
  V.~Bernard, N.~Kaiser and U.~G.~Mei\ss ner,
  {\it Low-energy theorems for weak pion production}, 
  Phys.\ Lett.\ B {\bf 331}, 137 (1994)
 % doi:10.1016/0370-2693(94)90954-7
  [hep-ph/9312307].
  %%CITATION = doi:10.1016/0370-2693(94)90954-7;%%
  %13 citations counted in INSpIRE as of 13 Feb 2019

\bibitem{Ando2000} 
%\cite{Ando:2000zw}
%\bibitem{Ando:2000zw} 
  S.~i.~Ando, F.~Myhrer and K.~Kubodera,
 {\it Capture rate and neutron helicity asymmetry for ordinary muon capture on hydrogen}, 
  Phys.\ Rev.\ C {\bf 63}, 015203 (2001)
  %doi:10.1103/physRevC.63.015203
  [nucl-th/0008003].
  %%CITATION = doi:10.1103/physRevC.63.015203;%%
  %28 citations counted in INSpIRE as of 13 Feb 2019

\bibitem{Yao2018} 
%\cite{Yao:2018pzc}
%\bibitem{Yao:2018pzc} 
  D.~L.~Yao, L.~Alvarez-Ruso, A.~N.~Hiller Blin and M.~J.~Vicente Vacas,
 {\it Weak pion production off the nucleon in covariant chiral perturbation theory}, 
  Phys.\ Rev.\ D {\bf 98}, 076004 (2018)
  %doi:10.1103/physRevD.98.076004
  [arXiv:1806.09364 [hep-ph]], \\
  %%CITATION = doi:10.1103/physRevD.98.076004;%%
  %2 citations counted in INSpIRE as of 13 Feb 2019 
  \pos{Yao},  contribution to this conference. 
  
  \bibitem{bkm95} 
  %\cite{Bernard:1995dp}
%\bibitem{Bernard:1995dp} 
  V.~Bernard, N.~Kaiser and U.~G.~Mei\ss ner,
  {\it Chiral dynamics in nucleons and nuclei}, 
  Int.\ J.\ Mod.\ Phys.\ E {\bf 4}, 193 (1995)
 % doi:10.1142/S0218301395000092
  [hep-ph/9501384].
  %%CITATION = doi:10.1142/S0218301395000092;%%
  %1039 citations counted in INSpIRE as of 14 Feb 2019
  
  
  \bibitem{hemmert} 
  %\cite{Hemmert:1996xg}
%\bibitem{Hemmert:1996xg} 
  T.~R.~Hemmert, B.~R.~Holstein and J.~Kambor,
  {\it Systematic 1/M expansion for spin 3/2 particles in baryon chiral perturbation theory}, 
  Phys.\ Lett.\ B {\bf 395}, 89 (1997)
 % doi:10.1016/S0370-2693(97)00049-X
  [hep-ph/9606456]. \\ 
  %%CITATION = doi:10.1016/S0370-2693(97)00049-X;%%
  %151 citations counted in INSpIRE as of 14 Feb 2019
  %
  %\cite{Hemmert:1997ye}
%\bibitem{Hemmert:1997ye} 
  T.~R.~Hemmert, B.~R.~Holstein and J.~Kambor,
  {\it Chiral Lagrangians and delta(1232) interactions: Formalism}, 
  J.\ Phys.\ G {\bf 24}, 1831 (1998)
%  doi:10.1088/0954-3899/24/10/003
  [hep-ph/9712496].
  %%CITATION = doi:10.1088/0954-3899/24/10/003;%%
  %288 citations counted in INSpIRE as of 14 Feb 2019 
  
  \bibitem{Pascalutsa} 
  %\cite{Pascalutsa:2002pi}
%\bibitem{Pascalutsa:2002pi} 
  V.~Pascalutsa and D.~R.~Phillips,
  {\it Effective theory of the delta(1232) in Compton scattering off the nucleon}, 
  Phys.\ Rev.\ C {\bf 67}, 055202 (2003)
  %doi:10.1103/PhysRevC.67.055202
  [nucl-th/0212024].
  %%CITATION = doi:10.1103/PhysRevC.67.055202;%%
  %187 citations counted in INSPIRE as of 21 Feb 2019
 
  
  \bibitem{ulf1} 
  %\cite{Bernard:1993ry}
%\bibitem{Bernard:1993ry} 
  V.~Bernard, N.~Kaiser, U.~G.~Mei\ss ner and A.~Schmidt,
  {\it Aspects of nucleon Compton scattering}, 
  Z.\ Phys.\ A {\bf 348}, 317 (1994)
  %doi:10.1007/BF01305891
  [hep-ph/9311354].
  %%CITATION = doi:10.1007/BF01305891;%%
  %102 citations counted in INSpIRE as of 15 Feb 2019 
  
  \bibitem{baru} 
  %\cite{Filin:2012za}
%\bibitem{Filin:2012za} 
  A.~A.~Filin, V.~Baru, E.~Epelbaum, H.~Krebs, C.~Hanhart, A.~E.~Kudryavtsev and F.~Myhrer,
  {\it Pion production in nucleon-nucleon collisions in chiral effective field theory: next-to-next-to-leading order contributions}, 
  Phys.\ Rev.\ C {\bf 85}, 054001 (2012)
 % doi:10.1103/PhysRevC.85.054001
  [arXiv:1201.4331 [nucl-th]]. \\ 
  %%CITATION = doi:10.1103/PhysRevC.85.054001;%%
  %15 citations counted in INSPIRE as of 15 Feb 2019
 %\cite{Baru:2013zpa}
%\bibitem{Baru:2013zpa} 
  V.~Baru, C.~Hanhart and F.~Myhrer,
  {\it Effective field theory calculations of $NN\to NN\pi$}, 
  Int.\ J.\ Mod.\ Phys.\ E {\bf 23}, %no. 4, 
  1430004 (2014)
  %doi:10.1142/S0218301314300045
  [arXiv:1310.3505 [nucl-th]].
  %%CITATION = doi:10.1142/S0218301314300045;%%
  %14 citations counted in INSPIRE as of 15 Feb 2019
 
  
  \bibitem{wirzba} 
  %\cite{Hanhart:2007mu}
%\bibitem{Hanhart:2007mu} 
  C.~Hanhart and A.~Wirzba,
  {\it Remarks on N N ---> N N $\pi$ beyond leading order}, 
  Phys.\ Lett.\ B {\bf 650}, 354 (2007)
  %doi:10.1016/j.physletb.2007.05.038
  [nucl-th/0703012 [nucl-th]].
  %%CITATION = doi:10.1016/j.physletb.2007.05.038;%%
  %23 citations counted in INSpIRE as of 15 Feb 2019

  
  \bibitem{sato}
  %\cite{Kim:2008qha}
%\bibitem{Kim:2008qha} 
  Y.~Kim, T.~Sato, F.~Myhrer and K.~Kubodera,
  {\it Two-pion-exchange and other higher-order contributions to the pp ---> pp $\pi^0$ reaction}, 
  Phys.\ Rev.\ C {\bf 80}, 015206 (2009)
 % doi:10.1103/physRevC.80.015206
  [arXiv:0810.2774 [nucl-th]].
  %%CITATION = doi:10.1103/physRevC.80.015206;%%
  %11 citations counted in INSpIRE as of 15 Feb 2019
 
  \bibitem{HNV} 
  %\cite{Hernandez:2007qq}
%\bibitem{Hernandez:2007qq} 
  E.~Hern\'{a}ndez, J.~Nieves and M.~Valverde,
  {\it Weak Pion Production off the Nucleon}, 
  Phys.\ Rev.\ D {\bf 76}, 033005 (2007)
 % doi:10.1103/PhysRevD.76.033005
  [hep-ph/0701149].
  %%CITATION = doi:10.1103/PhysRevD.76.033005;%%
  %174 citations counted in INSPIRE as of 21 Feb 2019 
  
  \bibitem{Watson}
  %\cite{Alvarez-Ruso:2015eva}
%\bibitem{Alvarez-Ruso:2015eva} 
  L.~Alvarez-Ruso, E.~Hern\'{a}ndez, J.~Nieves and M.~J.~Vicente Vacas,
  {\it Watson's theorem and the $N\Delta(1232)$ axial transition}, 
  Phys.\ Rev.\ D {\bf 93}, no. 1, 014016 (2016)
 % doi:10.1103/PhysRevD.93.014016
  [arXiv:1510.06266 [hep-ph]].
  %%CITATION = doi:10.1103/PhysRevD.93.014016;%%
  %18 citations counted in INSPIRE as of 22 Feb 2019
  
  \bibitem{HN}
%\cite{Hernandez:2016yfb}
%\bibitem{Hernandez:2016yfb} 
  E.~Hern\'{a}ndez and J.~Nieves,
  {\it Neutrino-induced one-pion production revisited: the $\nu_\mu n\to\mu^- n\pi^+$ channel}, 
  Phys.\ Rev.\ D {\bf 95}, no. 5, 053007 (2017)
%  doi:10.1103/PhysRevD.95.053007
  [arXiv:1612.02343 [hep-ph]].
  %%CITATION = doi:10.1103/PhysRevD.95.053007;%%
  %8 citations counted in INSPIRE as of 22 Feb 2019
  

\end{thebibliography}
\end{document}